 \documentclass[final,5p,times,twocolumn]{elsarticle}

\usepackage{amssymb}
\usepackage{multirow}
\usepackage{appendix}
\usepackage{verbatim}
\usepackage{mathtools}
\usepackage{float}
\usepackage{hyperref}
\usepackage{balance}
\usepackage{amsthm}

\journal{}

\begin{document}

%%\begin{linenumbers}
\begin{frontmatter}



\title{Design, characterization, and sensitivity of the supernova trigger system at Daya Bay}

\author[a]{Hanyu Wei\corref{why}}
\ead{weihy07@mails.tsinghua.edu.cn}
\author[a]{Logan Lebanowski}
\author[b]{Fei Li}
\author[a]{Zhe Wang}
\author[a]{Shaomin Chen}
\cortext[why]{Corresponding author. }

\address[a]{Department of Engineering Physics, Tsinghua University,
  Beijing, 100084, China}
\address[b]{Institute of High Energy Physics,
  Beijing, 100039, China}

\begin{abstract}
Providing an early warning of galactic supernova explosions from neutrino signals is important in studying supernova dynamics and neutrino physics. A dedicated supernova trigger system has been designed and installed in the data acquisition system at Daya Bay and integrated into the worldwide Supernova Early Warning System (SNEWS). Daya Bay's unique feature of eight identically-designed detectors deployed in three separate experimental halls makes the trigger system naturally robust against cosmogenic backgrounds, enabling a prompt analysis of online triggers and a tight control of the false-alert rate. The trigger system is estimated to be fully sensitive to 1987A-type supernova bursts throughout most of the Milky Way. The significant gain in sensitivity of the eight-detector configuration over a mass-equivalent single detector is also estimated. The experience of this online trigger system is applicable to future projects with spatially distributed detectors.
\end{abstract}


\end{frontmatter}

\section{Introduction}
\label{}
About two dozen supernova (SN) burst neutrinos were observed in the Kamiokande II, IMB, and Baksan experiments from stellar collapse SN 1987A when the star Sanduleak -69202 exploded in the Large Magellanic Cloud, about 50 kpc away from the Earth~\cite{Hirata1,Hirata2,Bionta,Bratton,Alekseev1,Alekseev2}. Besides the Sun, SN 1987A remains the only known astrophysical neutrino source that has provided a large range of physical limits on neutrinos as well as the core-collapse supernova mechanism~\cite{Raffelt,book,Lunardini,Masshierarchy}. Almost all of the gravitational binding energy of a stellar collapse is carried away by neutrinos and core-collapse supernovae are likely strong galactic sources of gravitational waves. Observations of both neutrinos and gravitational waves could provide deep insight into the core collapse of supernova explosions as well as other fundamental physics~\cite{Gravwave}.

Galactic SN explosions are rare, occurring with a rate of only a few per century~\cite{Rate}, so detecting neutrinos from a nearby SN is a once-in-a-lifetime opportunity. SN neutrinos are expected to arrive at the Earth a few hours before the visual SN explosion, which enables an early warning for a SN observation~\cite{Raffelt}. The Supernova Early Warning System (SNEWS)~\cite{snews,SNEWSpaper} collaborates with experiments sensitive to core collapse SN neutrinos, to provide the astronomical community with a very high-confidence early warning of a SN occurrence, pointing more powerful telescopes or facilities to the event.

The antineutrino detectors (ADs) of the Daya Bay experiment are designed to detect $\bar{\nu}_e$'s via the inverse beta-decay (IBD) interaction $\bar{\nu}_e+p\rightarrow e^++n$, with the primary goal of making a precision measurement the neutrino mixing angle $\theta_{13}$~\cite{dayabay1_1,dayabay1,dayabay2,dayabay3,dayabay4}. Each AD contains about 21.6~tons of liquid scintillator (LS) and 19.9~tons of liquid scintillator doped with gadolinium (Gd-LS), giving a total active target mass of $\sim$330 tons in 8 ADs. The water shields (about 2.3 kton in total) around the ADs collect too little light to efficiently detect SN burst neutrinos. A dedicated online supernova trigger system was installed in August 2013 and Daya Bay joined SNEWS in November 2014.

The experiments currently in SNEWS are Super-K, LVD, IceCube, Borexino, KamLAND, and Daya Bay~\cite{snews}. Though SN burst neutrinos come in all flavours in the few-tens-of-MeV range, the interaction rates in these experiments are dominated by IBD events~\cite{SNdetection}. Some main features~\cite{SNdetection,SNdetectors} of the experiments are summarized in Tab.~\ref{tab:SNEWS}.
\begin{center}
\begin{table}[!htb]
\centering
\footnotesize
\caption{\label{tab:SNEWS}Supernova neutrino detectors in SNEWS and their capabilities. N$_{\rm{IBD}}$ is the expected number of IBD events from a SN at 10 kpc, with an emission of $5\times10^{52}$ erg in $\bar{\nu}_e$'s, and an average $\bar{\nu}_e$ energy around 12 MeV, which is compatible with SN 1987A measurements.}
\begin{tabular*}{\columnwidth}[]{llllll}
\hline
Detector & Type & Location & Mass \scriptsize{(kt)} & N$_{\rm{IBD}}$ & $E_{\rm{th}}$ \scriptsize{(MeV)} \\
\hline
IceCube & *L.S. Ch. & Antarctic & 0.6/PMT & N/A & - \\
Super-K & Water Ch. & Japan & 32 & 7000 & 7.0 \\
LVD & Scint. & Italy & 1 & 300 & 4.0 \\
KamLAND & Scint. & Japan & 1 & 300 & 0.35 \\
Borexino & Scint. & Italy & 0.3 & 100 & 0.2 \\
Daya Bay & $\dagger$M.S. Scint. & China & 0.33 & 110 & 0.7 \\
\hline
\multicolumn{6}{l}{*~{Long-String Cherenkov}~~~$\dagger$~{Multiple-Site Scintillator}}
\end{tabular*}
\end{table}
\end{center}

Super-K is the only experiment with pointing capability, which is realized with neutrino-electron scattering interactions and thus applies to only a few percent of the total number of interactions~\cite{SK}. IceCube can detect a flux of MeV neutrinos via a collective increase in the rates of all PMTs caused by the Cherenkov light produced by the IBD positrons~\cite{AMANDA}. It cannot distinguish among neutrino flavors nor measure positron energy, although for a galactic SN, it can track the subtle features in the temporal development of the SN neutrino bursts \cite{IceCube}. KamLAND, Borexino, and Daya Bay have finer energy resolutions (e.g. $\sigma_E/E\approx3\%$ at 10 MeV for Daya Bay) and lower energy thresholds (below the IBD reaction threshold of 1.8 MeV), which provide sensitivity to the full spectrum of the SN burst electron-antineutrinos.

\begin{figure}[htb]
\centering
\includegraphics[width=\columnwidth]{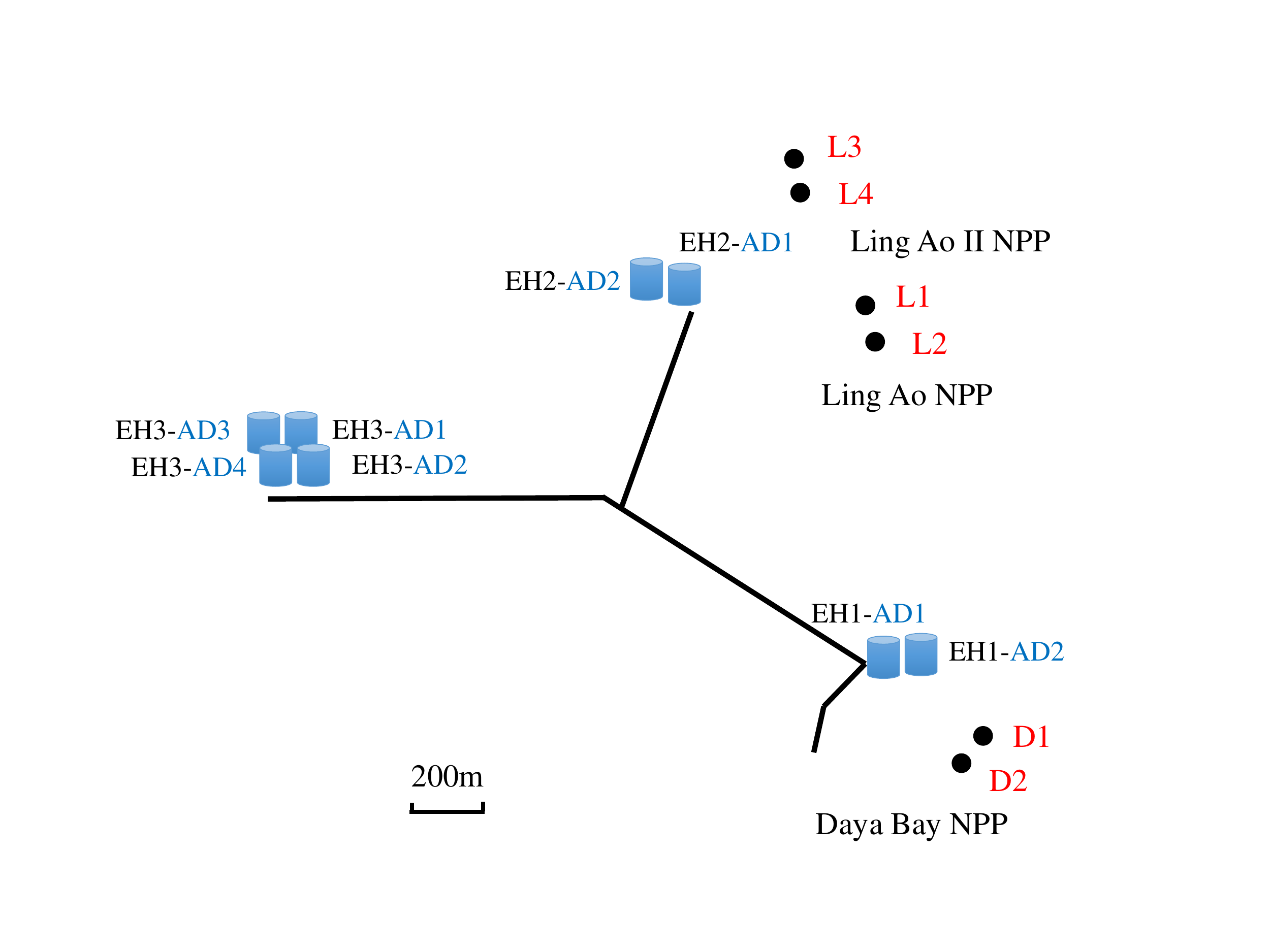}
\caption{Layout of the Daya Bay detectors. The dots represent reactor cores, labeled as D1, D2, L1, L2, L3 and L4. The black line represents horizontal tunnels that connect three underground experimental halls (EHs), where 8 ADs are installed.}
\label{fig:layout}
\end{figure}

Daya Bay has a unique feature of 8 identically-designed detectors deployed in three separate experimental halls (EHs, Fig.~\ref{fig:layout}), which are $>$~1~km apart from each other and whose maximum overburdens in equivalent meters of water are 250, 265, and 860, respectively. The online supernova trigger system at Daya Bay can provide an alert to the experiment and to SNEWS within 10 seconds, since the impact from muon-induced and accidental backgrounds or abnormal noise occurring in a single-detector is minimized by the separation of detectors. Daya Bay has the prominent features of a prompt alert and well controlled false-alert rate.

\section{Overview of the online supernova trigger system}
\label{}
\begin{figure*}[htb]
\centering
\includegraphics[width=0.8\paperwidth]{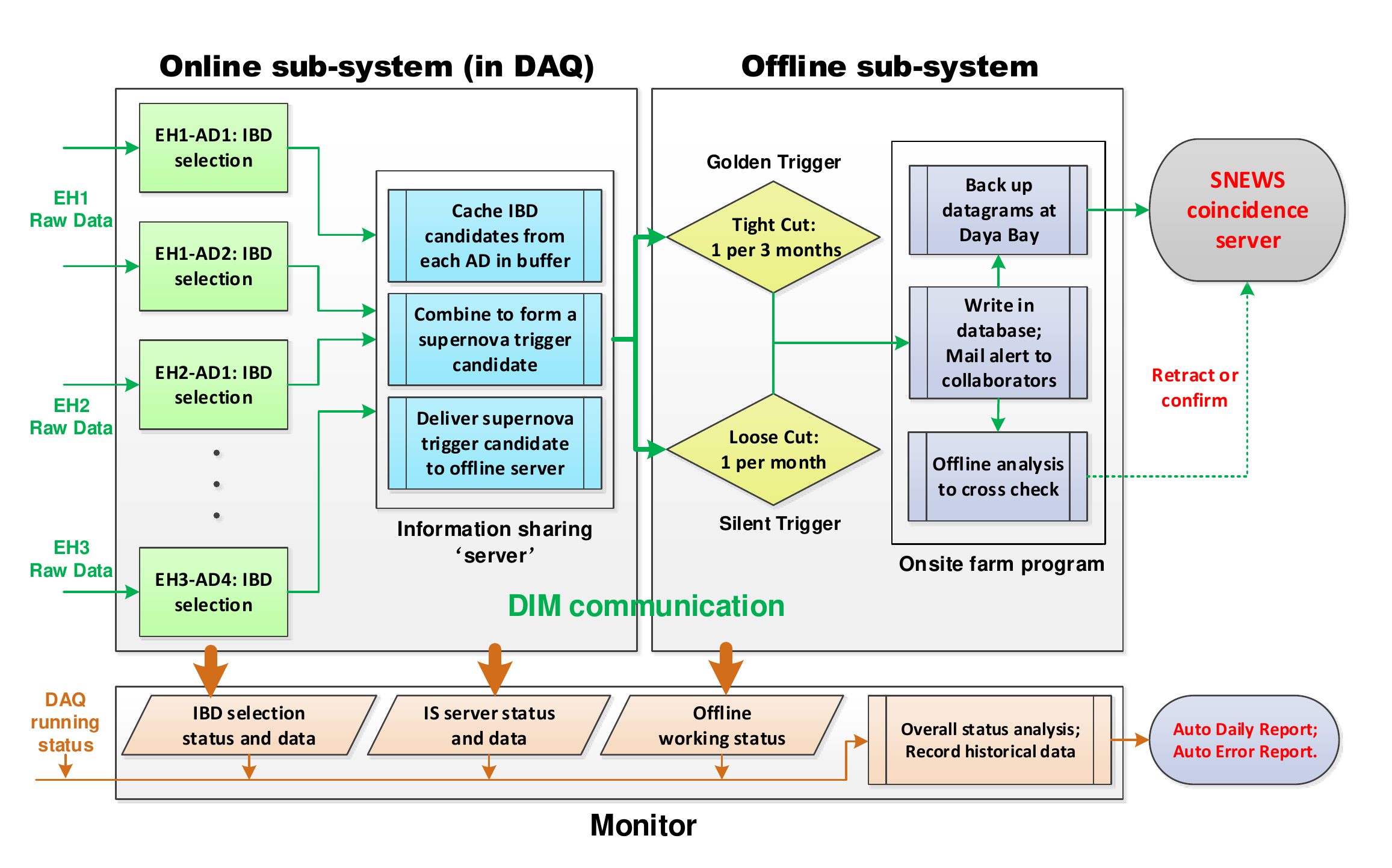}
\caption{Diagram of the online supernova trigger system of the Daya Bay experiment.}
\label{fig:snews_diagram}
\end{figure*}

An overview of the online supernova trigger system of Daya Bay is shown in Fig.~\ref{fig:snews_diagram}. The system is composed of three sub-systems: online, offline, and monitoring.

IBD events are reconstructed and selected (see Sec.~\ref{IBDevent}) in the online sub-system which contains software applications embedded in the Event Flow Distributer (EFD,~\ref{Abbrev}) of the Data Acquisition (DAQ) system~\cite{DAQ}. An IBD selection program for each AD accesses the raw data and provides the information of selected IBD candidates to an Information Sharing (IS,~\ref{Abbrev}) server. The IS server caches the IBD candidates from each AD to a 2-minute buffer and combines them every second to form an online supernova trigger candidate based on the previous 10-second window (see Sec.~\ref{triggerdecision}). Then, the online supernova trigger candidates are delivered to the offline sub-system via a Distributed Information Management (DIM,~\ref{Abbrev}) system, where they are compared against the trigger threshold.

The offline sub-system processes the input from the online sub-system with several standalone programs running in an onsite farm. Supernova candidate triggers are categorized as golden triggers (1 per 3 months) or silent triggers (1 per month). Both types are written into a database and generate an email alert to Daya Bay collaborators. The datagram is then sent to SNEWS. An offline cross check will be performed to confirm or retract the triggers. The datagram sent to SNEWS includes the experiment title, the alert date and time, the trigger duration, the number of neutrino signals and the type of the trigger.

Based on the existing DIM Name Server (DNS,~\ref{Abbrev}) in the Daya Bay Detector Control System (DCS) environment~\cite{DCS}, a real-time monitoring program located in the onsite farm communicates with the existing DAQ and the online and offline sub-systems of the supernova trigger system via DIM to obtain the trigger system status and trigger information. There is a 1 Hz heartbeat to the IBD selection program from the IS server. If the 8 ADs are not running simultaneously outside a tolerance of 2 minutes, a warning will be sent and the trigger system will continue to operate with the active ADs. Any abnormal running status for any level of the online supernova trigger system automatically generates an error report and mails it to experts. Real-time data is recorded, including the working hours of the online supernova trigger system, each AD's IBD candidate rate, unsolved errors, number of supernova triggers, and the network connection status to SNEWS. An automated daily report is sent to the supernova trigger working group, serving as a daily check of the online supernova trigger system.

\section{Algorithm of the online trigger}
\label{}
This section describes the online algorithm, which searches for a simultaneous increase in IBD rate in all ADs within a 10-second window. First, the IBD event selection is introduced, followed by the supernova trigger algorithm and the packing of consecutive supernova triggers. Lastly, the resulting detection probability is presented.

\subsection{IBD event selection}
\label{IBDevent}
\subsubsection{Event time, energy and vertex}
\label{recon}
In order to achieve a \textit{prompt} (fast) online supernova trigger, it was necessary for the event reconstruction to attain a balance between simplicity and effectiveness. The IBD prompt signal trigger time is identified as the time of the IBD event. Trigger times are provided by the GPS, which deviates from UTC time within 200 ns. Energy is reconstructed using an average PMT gain and an average energy scale (photoelectron yield per unit of deposited energy in the liquid scintillator) from calibration:
\begin{equation}
\label{recenergy}
  E~=~\frac{\rm{ADC~Sum}}{[\rm{Average~PMT~Gain}]\cdot [\rm{Average~Energy~Scale}]},
\end{equation}
where `ADC Sum' is the sum of ADC values with baselines subtracted for all PMT channels. The variation of the product of `Average PMT Gain' and `Average Energy Scale' calibration constants is less than 1\% per year. The ADC values are provided by the front-end electronics (FEE)~\cite{FEE}, which integrate each PMT signal. A charge-weighted method involving the PMT charges and PMT locations is utilized for a rapid vertex reconstruction, i.e.
\begin{equation}
\label{recvertex}
  \mathbf{X}=\frac{\sum_{\rm PMT} {\rm ADC}_{\rm PMT}\cdot \mathbf{X}_{\rm PMT}}{\sum_{\rm PMT} {\rm ADC}_{\rm PMT}}.
\end{equation}

The online reconstruction is sufficiently effective for online supernova triggering, though it cannot reach the same performance as the offline analysis reconstruction.

\subsubsection{IBD signal selection and background sources}
\label{signalselection}
Daya Bay ADs identify SN $\bar{\nu}_e$'s via the IBD reaction chain $\bar{\nu}_e+p\rightarrow e^++n$, $n+\rm{H/Gd}\rightarrow \rm{D/Gd}+\gamma/\gamma's$. The IBD selection basically follows the Daya Bay selection procedures described in previous publications~\cite{dayabay1_1,dayabay1,dayabay2,dayabay3,dayabay4}.

A simple PMT flasher cut is applied: max[ADC$_{\rm PMT}$]/[ADC Sum] $<$ 0.3. AD muon veto cuts are applied, where an AD (shower) muon event is defined to have a visible energy greater than 50 MeV (2.5 GeV) in an AD. The event is vetoed if its time since the previous AD (shower) muon is within 1 ms (1 s), which removes most of the muon spallation backgrounds and any follow-on triggers. Water-pool and RPC information are not accessible where the online trigger system is located.

The coincidence of the prompt signal from the positron with the delayed gamma emission of the neutron capture on Gd (nGd) or H (nH) provides a clear $\bar{\nu}_e$ signature against the background. Most of the neutrino energy is given to the positron. The delayed signal of an IBD event is either an 8 MeV $\gamma$'s cascade from nGd, or a 2.2 MeV $\gamma$ from nH. We set the online prompt energy threshold at 2 MeV for nGd events and 8 MeV for nH events, so that the majority of accidental backgrounds are removed~\cite{dayabay1_1, dayabay4}. The energy spectrum of SN burst $\bar{\nu}_e$ has an expected energy range up to $\sim$60 MeV with an average energy of 12-15 MeV~\cite{SNnuspec}. We cut prompt events above 50 MeV to preserve a good signal-to-background ratio against low energy cosmic-ray muons and cosmogenically-induced fast neutrons. The delayed signal energy cut is 6-10 MeV (1.9-2.8 MeV) for neutron capture on Gd (H). The prompt-delayed vertex distance is required to be less than 800 mm and
the time difference between the prompt (t$_{\rm p}$) and delayed (t$_{\rm d}$) candidate is required to satisfy 2~$<t_{\rm p}-t_{\rm d}<$~400~${\rm \mu}$s. The lower time limit suppresses coincidences due to electronic noise. The backgrounds are mainly reactor neutrinos in the lower prompt energy region and muon-induced fast neutrons in the higher prompt energy region according to Fig.~31 in~\cite{dayabay1_1} and Fig.~14 in~\cite{dayabay2}.

\subsubsection{SN $\bar{\nu}_e$ IBD selection efficiency of a single AD}
The selection efficiency of SN $\bar{\nu}_e$'s is estimated according to the nGd analysis~\cite{dayabay2} and nH analysis~\cite{dayabay4}~of the Daya Bay experiment.

We assume that the spectrum of supernova burst neutrinos follows a quasithermal distribution~\cite{SNnuspec}, for example,
\begin{equation}\label{spectrum}
  f_{\nu}(E)\propto E^{\alpha}e^{-(\alpha+1)E/E_{av}},
\end{equation}
where $E_{av}$ is the average energy and $\alpha$ a parameter describing the amount of spectral pinching. In the estimation of the efficiency of the prompt energy cut, $E_{av}$ is set to be 12.28~MeV and $\alpha$ to be 2.61, which correspond to the proto-neutron star cooling phase for SN $\bar{\nu}_e$ energy spectra.

The efficiency of the prompt energy cut was estimated to be $\sim$98\% for nGd events and $\sim$93\% for nH events. The product of the efficiencies of the other selection criteria can be estimated using efficiencies of the nGd analysis~\cite{dayabay2}, as well as the prompt energy cut efficiency and nH/nGd IBD candidate ratio of the nH analysis~\cite{dayabay4}. It is $\sim$89\% for nGd events in the GdLS volume, and $\sim$93\% and $\sim$56\% for nH events in the GdLS volume and LS volume, respectively. The total selection efficiency is $\sim$70\% and is used to determine the SN detection probability (Sec.~\ref{detectionprob}).

\subsection{Supernova Trigger}
\label{triggerdecision}
A supernova trigger is determined from a prompt analysis of IBD candidates from all ADs. We determine the rates of occurrence of various distributions of IBD candidates among ADs in a sliding 10-second window to set a false-alert rate. The duration of 10 seconds was chosen because it covers about 99\% of the luminosity carried off by all flavors of neutrinos and antineutrinos in a SN explosion~\cite{Raffelt}. The handling of overlapping supernova trigger determinations is also described in this section.
\subsubsection{Event combination and trigger determination}
\label{combine}
A trigger table is generated from all IBD candidate combinations for 8 ADs and sorted according to their rates of occurrence within a sliding 10-second window. Utilizing this table, a false-alert rate threshold is set, e.g. 1 per month or 1 per 3 months. A schematic of the trigger table is shown in Tab.~\ref{tab:triggertable}. For each row, the number under an AD is the number of IBD candidate events within a 10-second window in that AD and the last number is the corresponding rate of occurrence of the combination of events in that row. For example, the rate of occurrence of the combination 0-0-0-0-0-0-0-0 (no candidates in any of the 8 ADs) is r$_1$. Here $\sum_{i=1}^{\infty}$r$_i$ equals 1 (Hz) as the table enumerates all possible cases and the determination is made every second. The rate of occurrence is sorted in descending order, i.e. r$_i\geq$~r$_j$ when $i\leq j$ where $i$ and $j$ denote the row numbers of the table. Notice that r$_i$ is predicted based on the IBD candidate rate of each AD and the correlation between every two ADs in one experimental hall is taken into account. Elaboration of this prediction follows below.

For a single AD, the probability of the number of IBD candidates in a sliding 10-second window follows a Poisson distribution with the mean value -- 10~seconds~$\times~{\rm R^i_{IBD}}$, where R$^i_{\rm IBD}$ is the IBD candidate rate (Hz) of the $i$th AD. Different experimental halls are assumed to be mutually independent and the correlation coefficient between two ADs in the same hall is measured to be $<$ 0.01. The correlation originates from muon-induced backgrounds, mainly fast neutrons, that cause nearly simultaneous signals in detectors of the same experimental hall.

The rate of occurrence for each combination is predicted using several independent Poisson variables. Each AD has one Poisson variable for its own independent rate and every possible combination of two ADs in the same hall has one Poisson variable for their correlated rate. For instance, in the case of 2 ADs installed in one experimental hall (such as in EH1 and EH2), three independent Poisson variables are defined: $N_1, N_2$, and $N_s$. Suppose the event count of AD1 (AD2) in a 10-second window is $N_{\rm AD1}=N_1+N_s$ ($N_{\rm AD2}=N_2+N_s$), then the mean values for the three variables can be deduced from the measured mean values $\lambda(N_{\rm AD1})$ ($\lambda(N_{\rm AD2})$), i.e. 10~times~the~measured~${\rm R^{\rm AD1(AD2)}_{IBD}}$ [Hz], and the measured covariance $Cov(N_{\rm AD1},N_{\rm AD2})$:
\begin{equation}
\label{combination1}
\begin{array}{lll}
  \lambda(N_s) & = & Cov(N_s, N_s)~=~Cov(N_1+N_s, N_2+N_s) \\
  & = & Cov(N_{\rm AD1},N_{\rm AD2}),\\
  \lambda(N_1) & = & \lambda(N_{\rm AD1})~-~\lambda(N_s),\\
  \lambda(N_2) & = & \lambda(N_{\rm AD2})~-~\lambda(N_s).
\end{array}
\end{equation}
Therefore, each combination $(N_{\rm AD1},N_{\rm AD2})$ could arise from numerous possible cases of $(N_1, N_2, N_s)$ and its probability obtained by summing over the possible cases:
\begin{equation}
\label{combination2}
\begin{array}{rll}
    & P(N_{\rm AD1},N_{\rm AD2}) \\
  = & \sum_{i,~j,~k} \{ ~{\rm Po}(\lambda(N_s), i) \cdot {\rm Po}(\lambda(N_1), j) \cdot {\rm Po}(\lambda(N_2), k)~ \} \,
\end{array}
\end{equation}
where $i, j, k$ are non-negative integers and $i + j = N_{\rm AD1}$, $i + k = N_{\rm AD2}$ and Po is the Poisson distribution function. For the scenario of 4 ADs in an experimental hall (such as in EH3), 4~+~6 independent Poisson variables are defined and the details of the calculation are analogous to the 2-AD scenario.

\begin{center}
\begin{table}[!htb]
\centering
\caption{\label{tab:triggertable}Schematic of the trigger table for online supernova trigger determination. EH1-AD1 to EH3-AD4 indicates the 8 ADs in the three experimental halls at Daya Bay. The table is sorted in descending order of the ``Rate'' of occurrence.}
\begin{tabular}{cccc}
\hline
EH1       & EH2        & EH3                  & Rate (Hz)  \\
\small{AD1~AD2} & \small{AD1~AD2} & \small{AD1~AD2~AD3~AD4} & (r$_i>$ r$_{i+1}$) \\
\hline
0 ~~ 0    & 0 ~~ 0     & 0 ~~ 0 ~~ 0 ~~ 0     &  r$_1$ \\
0 ~~ 1    & 0 ~~ 0     & 0 ~~ 0 ~~ 0 ~~ 0     &  r$_2$ \\
\vdots    & \vdots     & \vdots               & \vdots \\
0 ~~ 0    & 0 ~~ 1     & 0 ~~ 1 ~~ 0 ~~ 0     &  r$_n$\\
2 ~~ 0    & 0 ~~ 1     & 0 ~~ 0 ~~ 0 ~~ 0     &  r$_{n+1}$\\
\vdots    & \vdots     & \vdots               & \vdots \\
\hline
\end{tabular}
\end{table}
\end{center}

\begin{figure}[htb]
\centering
\includegraphics[width=\columnwidth]{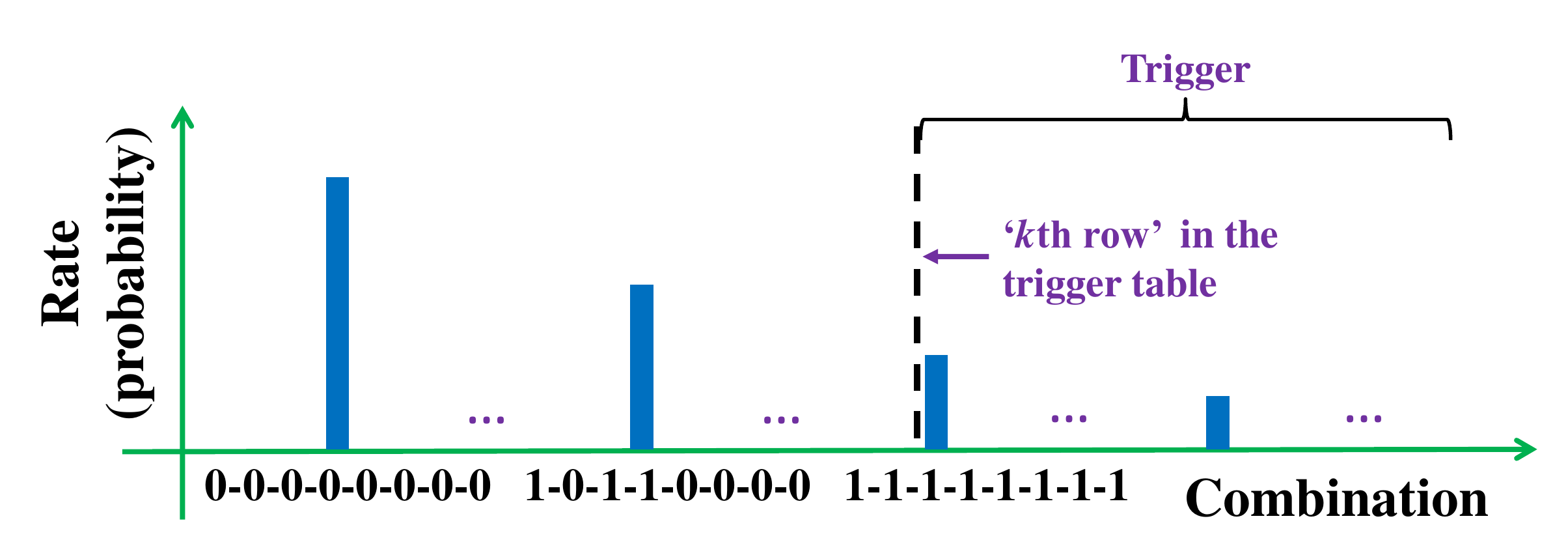}
\caption{Determination of the trigger cut from the trigger table. All possible combinations are enumerated in descending order of their rates of occurrence. The dashed line is determined according to a required false-alert rate threshold. On the right side of the dashed line, the sum of the rates of occurrence for the combinations is the false-alert rate.}
\label{fig:tableillustration}
\end{figure}

The trigger table is used to control the false-alert rate due to backgrounds. For a given false-alert rate threshold (e.g. 1/month) $P_{\rm{DYB}}$ [Hz], we can find the $k$th row in the table that satisfies
\begin{equation}
\label{equ:threshold}
  \sum_{i=k+1}^{\infty}{\rm r}_i\leq P_{\rm{DYB}}~~{\rm and}~~\sum_{i=k}^{\infty}{\rm r}_i\geq P_{\rm{DYB}}.
\end{equation}
The combinations below the $k$th row are supposed to be online supernova triggers, in which case the false-alert rate is about $P_{\rm{DYB}}$ Hz. An illustration of the trigger table with a trigger cut is shown in Fig.~\ref{fig:tableillustration}.

To suppress unexpected triggers from non-astrophysical correlated bursts (e.g.~electronic noise) in one detector or one experimental hall, a uniformity cut based on the $\chi^2$ method was applied with negligible loss of sensitivity to supernova explosions. This $\chi^2$ is the minimum value of $\sum_{i=1}^8\frac{(N_i-\lambda)^2}{\lambda}$ where $N_i$ is the number of IBD candidates of AD$_i$ and $\lambda$ is the average of $N_i$, which assumes SN $\bar{\nu}_e$ events to be distributed uniformly among the ADs. The cut is set to the 99\% confidence level. An analogous uniformity cut (95\% confidence level) for the three experimental halls was also applied for the case of an abnormal event cluster in a single experimental hall.

\subsubsection{Packing consecutive supernova triggers}
\label{packing}
\begin{figure}[htb]
\centering
\includegraphics[width=\columnwidth]{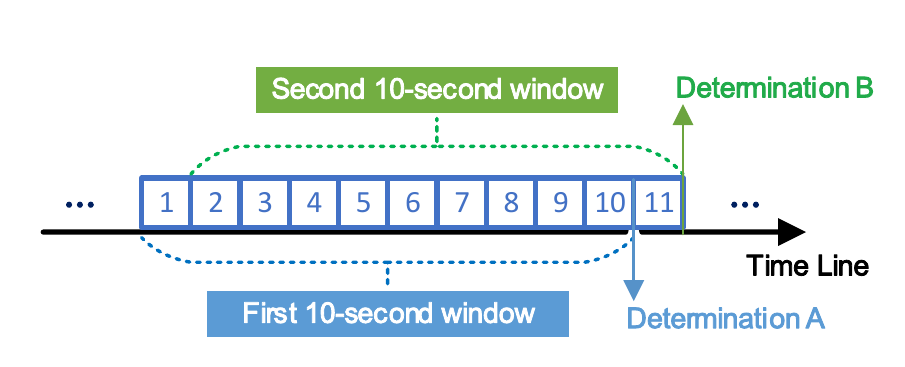}
\caption{Demonstration of two overlapping 10-second windows of two consecutive supernova trigger determinations.}
\label{fig:packing}
\end{figure}

As illustrated in Fig.~\ref{fig:packing}, online supernova trigger determinations, e.g. A and B, are made second-by-second and based on their previous 10-second window. Due to the correlation of overlapping windows, it is likely that a series of consecutive triggers would occur and be sent to Daya Bay collaborators or SNEWS with redundant information. To avoid this, we pack such triggers as a single supernova trigger and send one datagram to SNEWS that includes the entire supernova trigger duration and the total number of IBD candidates in all triggered 10-second windows.

The false-alert rate $P_{\rm{SNEWS}}$ [Hz] required by SNEWS must be converted to a Daya Bay trigger threshold $P_{\rm{DYB}}$ [Hz], which is defined in Eq.~\ref{equ:threshold}. The relation between $P_{\rm{SNEWS}}$ and $P_{\rm{DYB}}$ is expressed as:
\begin{equation}
\label{equ:packing}
  P_{\rm{SNEWS}}~=~\frac{P_{\rm{DYB}}}{\bar{N}},
\end{equation}
where $\bar{N}$ is the average number of consecutive online supernova triggers. Because $\bar{N}$ is difficult to calculate analytically and computationally expensive to simulate numerically, an upper limit was estimated (see~\ref{app:packing}).

For the final result, we take $\bar{N}=$3 based on the upper limit 5.5 and the lower limit 1. For about 1 year of operation, this implementation of packing has worked as intended. If needed, the value of $\bar{N}$ can be tuned based on more observations in the future.

\subsection{Detection probability}
\label{detectionprob}
With the SN explosion distance to the Earth ($D$) and the luminosity of electron-antineutrino emission ($L_{\bar{\nu}_e}$), a single AD's expected SN $\bar{\nu}_e$ event number can be determined by~\cite{Raffelt}
\begin{equation}
\label{SNevent}
  N_{AD} = N_0\times\frac{L_{\bar{\nu}_e}}{5\times10^{52}erg}\times(\frac{10kpc}{D})^2,
\end{equation}
where $N_0$ is the expected number of SN $\bar{\nu}_e$ events in a 10-second window corresponding to $\sim$0.33 ktons of liquid scintillator, a $5\times10^{52}$~erg luminosity, a $10$~kpc distance, and Daya Bay's selection efficiency of $\sim$70\%. The nominal value of $N_0 $ is $\sim$10.

Based on an expected number of SN signals and the background rates, the detection probability of a supernova explosion can be calculated by summing the probabilities of the combinations (SN signals plus backgrounds) surviving the trigger cut. In this calculation, the number of SN signals is assumed an independent Poisson variable for each AD, and the IBD candidate (background) rates, i.e. R$^i_{\rm IBD}$ (Sec.~\ref{combine}) for all ADs, are needed to determine the probabilities of the combinations and to set the trigger cut. The IBD candidates are mainly reactor neutrinos associated with nGd IBDs of lower prompt energy (reactor nH IBDs are suppressed with the prompt energy cut) and muon-induced fast neutrons of higher prompt energy~\cite{dayabay1_1, dayabay4}.

\begin{figure}[htb]
\centering
\includegraphics[width=\columnwidth]{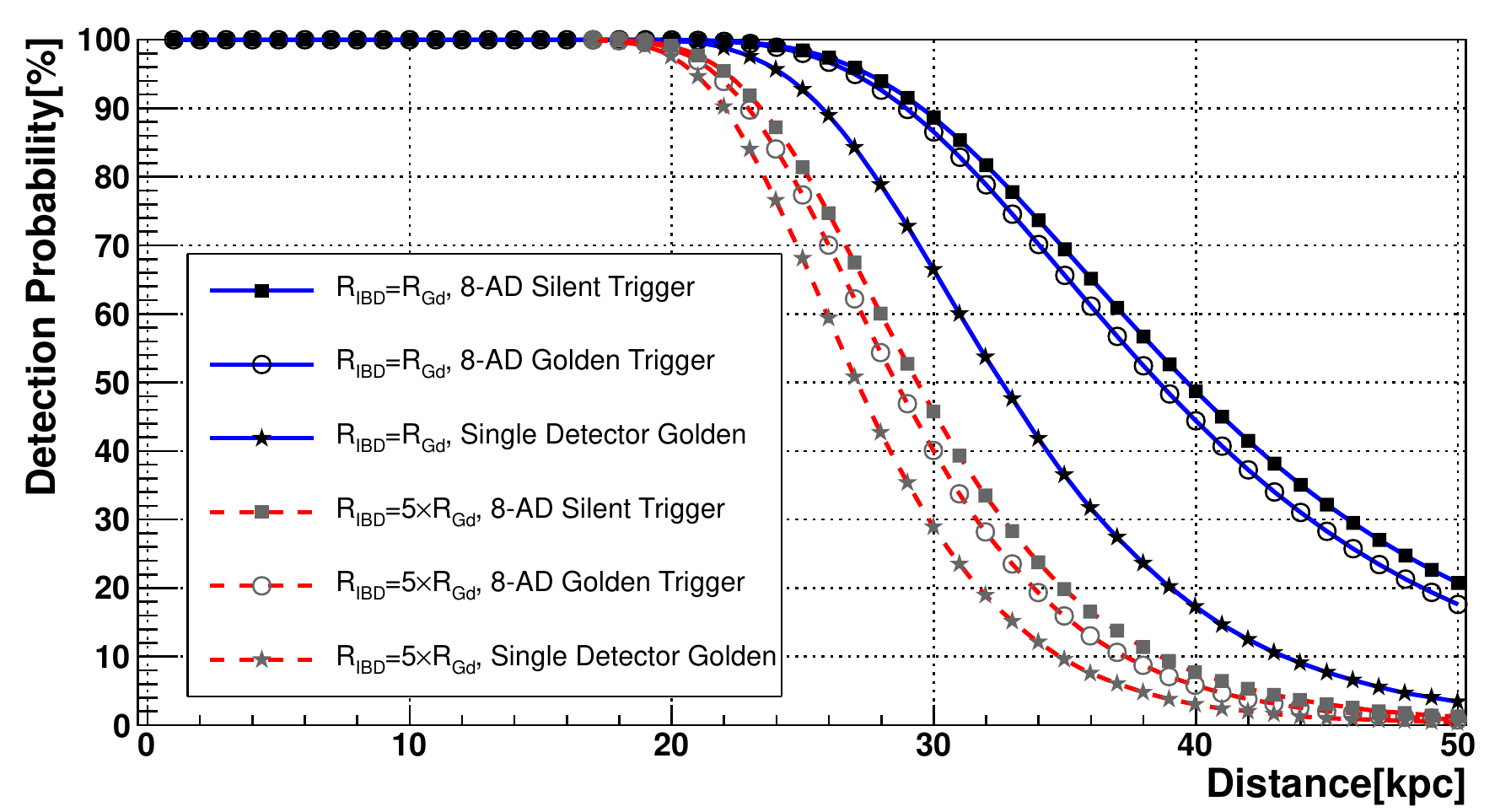}
\caption{The X-axis is SN explosion distance to the Earth and the Y-axis is the corresponding detection probability. ``8-AD golden trigger'' corresponds to the result with false-alert rate $<$1/3months, and ``8-AD silent trigger'' corresponds to that with false-alert rate $<$1/month. ``Single Detector Golden'' is the scenario in which the 8-AD target mass is combined into a single detector, also with a false-alert rate $<$1/3months. Two different values of R$_{\rm IBD}$ are assumed for solid blue curves and dashed red curves. }
\label{fig:detectionprobability}
\end{figure}

The detection probability is demonstrated for two cases: R$^i_{\rm IBD}$ = R$^i_{\rm Gd}$ and R$^i_{\rm IBD}$ = $5\times$R$^i_{\rm Gd}$, where R$^i_{\rm Gd}$ is the nGd IBD candidate rate of the $i$th AD, which is estimated from~\cite{dayabay2}. We calculate the detection probability of the SN explosion as a function of distance to the Earth. The result is shown in Fig.~\ref{fig:detectionprobability}.
When R$^i_{\rm IBD} < 5\times$R$^i_{\rm Gd}$, the golden trigger has a greater than 90\% probability for the most distant edge of the Milky Way, 23.5 kpc from the Earth. The silent trigger may have an additional 5\% probability to detect SN explosions. The ``Single Detector Golden'' curves represent the golden trigger detection probability if the target masses of all 8ADs were combined into a single detector and the background rate was the average of all the Daya Bay ADs. Comparison of ``8-AD'' curves and ``Single Detector'' curves illustrates the significant gain in sensitivity of multiple detectors over a mass-equivalent single detector.

\section{Test and characterization}
\label{}
Various tests were carried out at Daya Bay before formally joining SNEWS. The tests showed that the entire design of the system is effective. Particularly, the false-alert control was proven to be effective with tests of various trigger cuts. Moreover, the extra workload and dead time brought to the DAQ system are negligible. It was found that the system can handle IBD event input at 1 kHz, though it could handle 26 kHz in theory. Some characterization tests also show that the time latency of the online supernova trigger system is less than 10 seconds, indicating a quick processing of the raw data.

Communication tests with SNEWS began in December 2013 and after some compiling and debugging of the specific package for communication, the ``real test'' started in February 2014 for which the trigger threshold was set and everything configured according to SNEWS's requests. So far the online supernova trigger system has been running stably and effectively, watching for any increases in multi-AD signals within sliding 10-second windows. The trigger system is operating during only DAQ physics runs with a 100\% live time. The numbers (rates) of silent triggers and golden triggers are 4 (0.67 per month) and 2 (1 per 3 months), respectively, since November 2014 when Daya Bay officially joined SNEWS~\cite{snews}. All these triggers were identified as backgrounds and ruled out by offline checks.

\section{Conclusion}
\label{last}
We have developed and implemented an online supernova trigger system for the Daya Bay Reactor Neutrino Experiment based on the running DAQ system. The feature of Daya Bay's multiple ADs distantly-deployed allows a rapid trigger algorithm that tightly controls the false-alert rate and provides a significant gain in sensitivity over a single detector of equivalent mass. This online supernova trigger system is running smoothly with a $\sim$96\% live time and in cooperation with SNEWS. It is estimated to be fully sensitive to 1987A-type supernova bursts throughout most of the Milky Way.

Further optimizations may be applied to this online trigger system. The experience from this work could benefit an offline supernova study in Daya Bay as well as other experiments or facilities with multiple detectors who would develop an online supernova trigger. One such application could be to the detectors that will be distributed in the China JinPing Underground Laboratory~\cite{CJPL} where the deep rock cover may allow only small neutrino detectors to be installed. Supernova search results from the Daya Bay experiment including the online triggers and offline analysis are expected to be published in the future.

\section*{Acknowledgments}
This work is supported in part by the Ministry of Science and Technology of China, the Chinese Academy of Sciences, the National Natural Science Foundation of China (No. 11235006), the Key Laboratory of Particle \& Radiation Imaging (Tsinghua University) and the Tsinghua University Initiative Scientific Research Program. We gratefully acknowledge Kate Scholberg and our Daya Bay collaborators for their advice and assistance.



\appendix
\section{Abbreviations}
\label{Abbrev}
\begin{itemize}
  \item EFD: Event flow distributer of the Daya Bay DAQ, can receive data from each detector. The EFD will send data and fill data quality monitoring histograms, then publish the data to an information sharing server~\cite{DAQ}. The online supernova trigger system utilizes this function to access the raw data, reconstruct the useful information and make the IBD selection.
  \item IS: Information sharing, one of the online software packages of the Daya Bay DAQ, contains classes to support information sharing in the DAQ system. It can report error messages, to publish states and statistics, to distribute histograms built by the sub-systems of the DAQ system and detectors, and to distribute events sampled from different parts of the experiment's data flow chain~\cite{DAQ}. The online supernova trigger system uses the existing IS server to receive the IBD candidates from each AD and combine them to form an online supernova trigger candidate.
  \item DIM~\cite{DIM}: Distributed information management system, a communication system for distributed environments. It is a portable, light weight package for information publishing, data transfer and inter-process communication. Like most communication systems, DIM is based on the client/server paradigm. Servers ``publish'' their services by registering them with the name server. Clients ``subscribe'' to services by asking the name server which provides the service and then contacts the server directly. Fig.~\ref{fig:DIM} shows how DIM components interact.
\begin{figure}[htb]
\centering
\includegraphics[width=0.8\columnwidth]{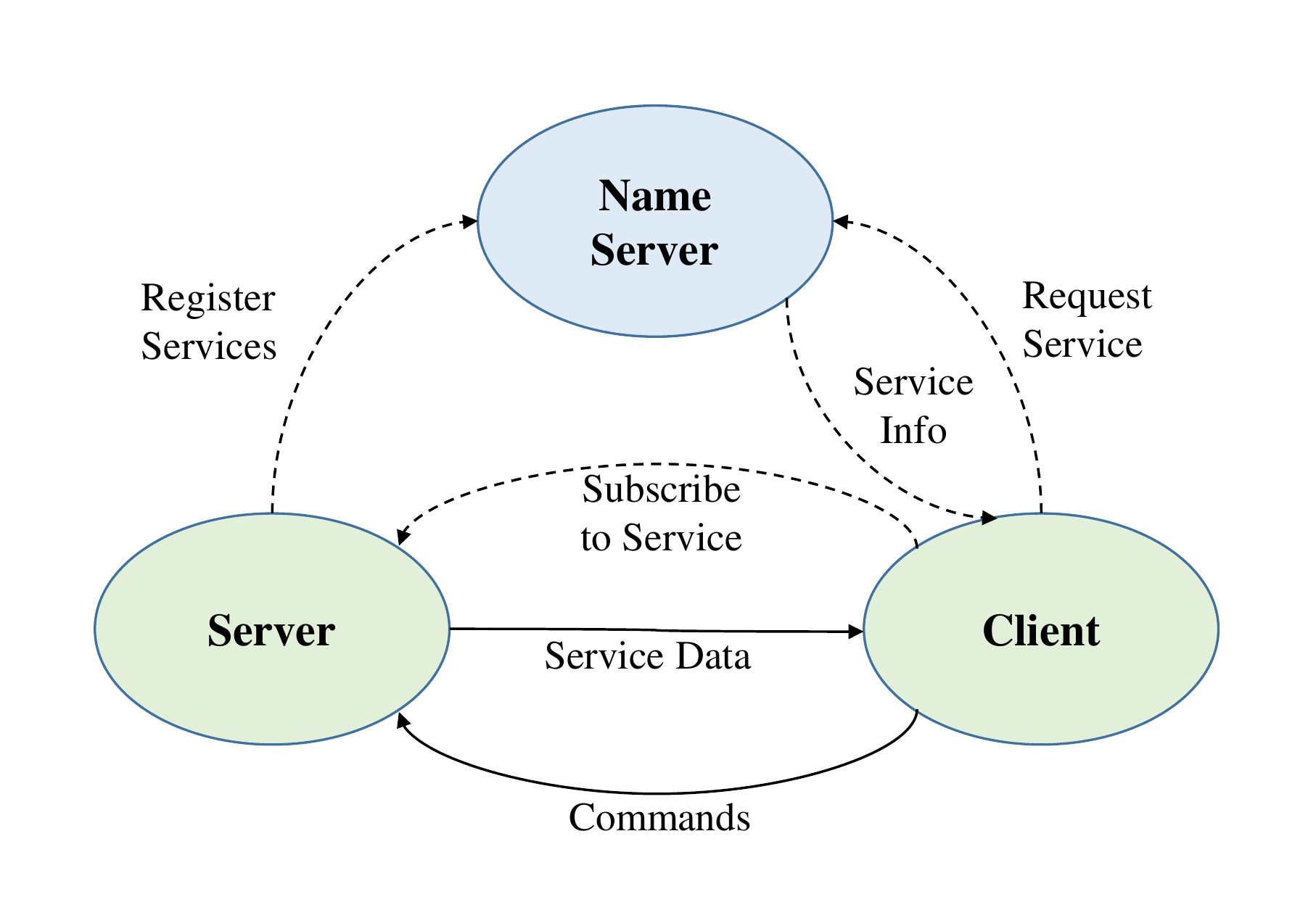}
\caption{Schematic DIM component diagram.~\cite{DIM}}
\label{fig:DIM}
\end{figure}
      The online supernova trigger system implements the data transfer, information publishing and communication between different parts through DIM.
  \item DNS~\cite{DIM}: DIM name server. The basis in the DIM is the concept of ``service''. A service is normally a set of data and it is recognized by a name - ``named services''. In order to allow for network transparency as well as easy recovery from crashes and migration of servers, a name server was introduced (dim name server). It keeps an up-to-date directory of all the servers and services available in the system. Any newly-built client can directly request the service from the existing DNS.
\end{itemize}

\section{Upper limit of the average number of consecutive supernova triggers}
\label{app:packing}
In Sec.~\ref{packing}, $\bar{N}$ is defined as the average number of consecutive supernova triggers, which relates the false-alert rate required by SNEWS with the Daya Bay trigger threshold for trigger determination (Eq.~\ref{equ:packing}).

We assume a sequence of consecutive triggers and use $A_i$ to indicate that a trigger is issued for the $i$th 10-second window and $\bar{A}_i$ to indicate no trigger. The probability of a trigger in the $i$th 10-second window is thus $P(A_i)$, equal to $P_{\rm{DYB}}$. $\bar{N}$ is expressed in terms of the conditional probabilities:
\begin{equation}
\label{equ:aveN}
\begin{array}{rcl}
\bar{N} & = & 1\cdot P(\bar{A}_2|A_1) + 2\cdot P(A_2\bar{A}_3|A_1) + ... + \\
&   & i\cdot P(A_2\cdots A_i\bar{A}_{i+1}|A_1) + ... \\
& = & 1\cdot P_1 + 2\cdot P_2 + ... + i\cdot P_i + ... ,
\end{array}
\end{equation}
where $P_i=P(A_1\cdots A_i\bar{A}_{i+1})/P(A_1)$. Since $A_i$ with $i~>~10$ is independent from $A_1$ and $P(A_i)=P_{\rm{DYB}}\ll1$ [Hz] (e.g. 1/month), terms containing $A_i$ with $i>10$ are not taken into account in the calculation below. By symmetry, $P_1=P(A_1\bar{A}_2)/P(A_1)=P(A_2\bar{A}_3)/P(A_1)$. $P(A_2\bar{A}_3)/P(A_1)>P(A_1A_2\bar{A}_3)/P(A_1)=P_2$, due to the occurrence of $A_1$ in the numerator. So $P_1>P_2$ is established. Similarly, we get $P_{i}>P_{i+1}$. Because
\begin{equation}
  \sum_{i=1}^{10}P_i<1~{\rm and}~P_1>P_2>\cdots >P_{10},
\end{equation}
$\bar{N}$ can be constrained to an upper limit by assuming $\sum_{i=1}^{10}P_i=$1 and $P_1=P_2=\cdots =P_{10}=$~0.1:
\begin{equation}
\label{equ:aveN2}
\begin{array}{ll}
\bar{N} & \cong 1\cdot P_1 + 2\cdot P_2 + ... + 10\cdot P_{10}  \\
& < 1\cdot 0.1 + 2\cdot 0.1 + 3\cdot 0.1 + ... + 10\cdot 0.1~=~5.5.
\end{array}
\end{equation}
For the real case, some $P_i$'s with larger $i$ should be less than 0.1 and the rest of $P_i$'s with smaller $i$ must be equal to or greater than 0.1, leading to a smaller $\bar{N}$ than in Eq.~\ref{equ:aveN2}.

%%\end{linenumbers}


\balance

\end{document}